\documentclass[preprint,aps,amsmath,pre,showpacs]{revtex4-1}
\usepackage[utf8]{inputenc} 
\usepackage{graphicx}
\usepackage{amsmath,amssymb,amsthm,amsfonts}
\usepackage{epsfig}
\usepackage{color}
\usepackage{xcolor}
\usepackage{float}
\usepackage{ulem}

\definecolor{brandeisblue}{rgb}{0.0, 0.44, 1.0}
\newcommand\bluesout{\bgroup\markoverwith{\textcolor{brandeisblue}{\rule[0.5ex]{2pt}{1.5pt}}}\ULon}
\newcommand\redsout{\bgroup\markoverwith{\textcolor{red}{\rule[0.5ex]{2pt}{1.5pt}}}\ULon} 
\begin{document}

\title{Modular structure in \textit{C. elegans} neural network and its response to external localized stimuli}

\author{Carolina A. Moreira and Marcus A.M. de Aguiar}
\email{corresponding author:aguiar@ifi.unicamp.br}

\affiliation{Instituto de Física F\'{\i}sica `Gleb Wataghin',
Universidade Estadual de Campinas, Unicamp\\ 13083-970, Campinas,
SP, Brazil}

\begin{abstract}

Synchronization plays a key role in information processing in neuronal networks. Response of specific groups of neurons
are triggered by external stimuli, such as visual, tactile or olfactory inputs. Neurons, however, can be divided into
several categories, such as by physical location,  functional role or topological clustering properties. Here we study
the response of the electric junction \textit{C. elegans} network to external stimuli using the partially
forced Kuramoto model and applying the force to  specific  groups of neurons. Stimuli were applied to
topological modules, obtained by the ModuLand procedure, to a ganglion, specified by its anatomical localization, and to
the functional group composed of all sensory neurons. We found that topological modules do not contain purely
anatomical groups or functional classes, corroborating previous results, and that stimulating different classes of
neurons lead to very different responses, measured in terms of synchronization and phase velocity correlations. In all
cases, however, the modular structure hindered full synchronization, protecting the system from seizures. More
importantly, the responses to stimuli applied to topological and functional modules showed pronounced patterns of
correlation or anti-correlation with other  modules that were not observed when the stimulus was applied to ganglia.

\end{abstract}

%05.45.Xt    Synchronization; coupled oscillators
%89.75.-k    Complex systems

%keywords    synchronization;networks;Kuramoto model;periodic force

\maketitle

\section{Introduction}
\label{intro}

Understanding the network of neuronal connections in the brain is key to unravel the way it works and processes
information.  The complexity of these networks has been emphasized by many authors \cite{Sporns2013}, and characterized
with different measures, such as degree distribution, transitivity and betweenness centrality \cite{Rubinov2010}. An
important feature of neural networks is their high degree of heterogeneity, in the sense that the number of connections
per neuron varies considerably and typically displays some sort of power law distribution. Moreover, neurons tend to
form communities, where the density of connections is higher within than among communities. Because connections are
constrained by anatomical features, neurons are also organized into physically arranged clusters, such as lobes or
ganglia, where neurons with different functional roles coexist \cite{sporns-betzel2016,bacik2016,antono2015}. 

Communities are often related to specialized areas of the brain and their number and structure are an indication of how
many different tasks it can perform  \cite{kim2014}. The integration of communities, on the other hand, measures how
well the outcomes of these different processes can combined to build a global view of the inputs
\cite{sporns-betzel2016}. When triggered by external stimuli, such as visual or olfactory inputs, the information
processing occurs by the synchronized firing of neurons responsible to process those specific tasks \cite{Schmidt2015,
	gray1994}. Synchronization of larger sets of neurons, or even global synchronization, indicates cerebral disorders
\cite{Uhlhaas2006} such as epilepsy \cite{global2012} and Alzheimer's disease \cite{global2016}, causing a general
breakdown in the neuronal network. Lack of synchronization, on the other hand, suggests difficulty to respond to the
stimulus or to function properly, as reported in unsuccessful overnight memory consolidation in old people
\cite{helfrich2018}, deficiency in the auditory-motor connections \cite{sowinskia2013} or brain disorders in autistic
individuals \cite {autism2013, autism2011}. In this context, the knowledge of the organization of different types of
neurons in the network and their segregation into modules or communities is fundamental to understand how stimuli affect
the target module and under what conditions it propagates to other regions leading to global or poor responses.

In this work we probe the community structure of the neural electrical junction network of the \textit{C. elegans} using
the partially forced Kuramoto model of synchronization \cite{us}. We aim to understand how the network responds to
external localized stimuli and which modules are more affected when a specific group of neurons, that can be a
functional group or a physically arranged module, is stimulated. We use two different metrics to characterize the
overall behavior of the network under a localized stimulus: the synchronization of neurons within and between modules,
as measured by the usual Kuramoto order parameter, and the phase-velocity inter-neuron correlation. We want to
	investigate the behavior of the system as a function of parameters such as stimulus intensity and inter-neuron
	connection strength. In particular we are interested in cases leading to global induced synchronization and
	highly correlated behavior, where the network responds as a whole, or to uncorrelated states, where neurons do not react
	to each other. Our simulations are guided by the results of a previous paper \cite{us} where we studied the
partially forced Kuramoto model on synthetic networks, using the external force to simulate a
localized stimulus.

\textit{C. elegans} is a nematode animal, unsegmented and with bilateral symmetry, exhibiting physiological similarity
to mammals as regards the nerves and neurotransmiters morphologies and it is considered a model organism in studies of
disorders related to human nervous system, such as epilepsy \cite{epi1,epi2} and Parkinson's disease \cite{parkinson1,
	parkinson2}. It was the first multicelular animal to be its whole nervous system mapped, containing only 302 neurons.
Its neural network has a huge data with open source, such as the WormAtlas \cite{wormatlas} and the OpenWorm
\cite{openworm}. Because it is a real complex network, we choose this small neural network, extracting all necessary
data from WormAtlas.

The 248 neurons of the electrical junction network are anatomically classified as belonging to head, body or tail, and
neuron types are divided into motoneurons, interneurons and sensory neurons. We have also performed a classification
into 10 ganglia (A: anterior ganglion, B: dorsal ganglion, C: lateral ganglion, D: ventral ganglion, E: retrovesicular
ganglion, F: posterolateral ganglion, G: ventral cord neuron group, H: pre-anal ganglion, J: dorsorectal ganglion, K:
lumbar ganglion \cite{wormatlas}) which is a finner division of the anatomical one.

The network is first decomposed into three modules based on topological properties and numbered by 1, 2 and 3 from
largest to smallest. This modularization procedure was made on software Cytoscape using the app ModuLand
\cite{moduland}. Each module contains neurons from the three anatomical parts, and consequently the 10 ganglia, and of
the three types. We applied the stimulus to the largest module, then on the ganglion C and finally to the sensory
neurons and then we observe the response of other neurons. We show that no single partition of the brain into
communities can account for its behavior under stimuli. All partitions analyzed here, topological, anatomical and
functional, play a role in the response to external localized stimuli, revealing the complexity of the brain's wiring
and function.

This paper is organized as follows: in section II we describe the materials and methods, showing the partially forced
Kuramoto model, the \textit{C. elegans} neural connectome and the order parameters used to measure the state of the
network. The results of numerical calculations and its analysis are in section III. Finally, we summarize our discussion
in section IV.

% % % % % % % % % % % % % % % % % % % % % % % % % % % % % % % % % % %
% % % % % % % % % % % % % % % % % % % % % % % % % % % % % % % % % % %
\newpage
\section{Materials and Methods}
\label{mm}

\subsection{Partially forced Kuramoto model}
\label{pfk}

The Kuramoto  model of coupled oscillators \cite{Kuramoto1975} is a paradigm in the study of
synchronization and has been explored in connection with biological systems, neural networks and the
social sciences \cite{Rodrigues2016, Acebron2005}. Here we consider a modified version of the
original Kuramoto model where each oscillator interacts only with a subset of the other oscillators,
as specified by a network of connections \cite{Arenas2008}. Moreover,  part of the oscillators also
interacts with an external periodic force \cite{Sakaguchi1988,Ott2008,Childs2008,us}. The force can
be interpreted as an external stimulus and the set of oscillators coupled to it represents the
'interface' of the system, like the photo-receptor cells in the eye \cite{gray1994}. The oscillators are
described by their phase $\theta$ and system is governed by the equations \cite{us} 
\begin{equation}
\dot{\theta_i} = \omega_i + F \, \delta_{i,C} \sin(\sigma t - \theta_i) + 
\frac{\lambda}{k_i} \sum_{j=1}^{N} A_{ij} \sin (\theta_j - \theta_i) ,
\label{forced1}
\end{equation}
where $N$ is the number of oscillators, $\lambda$ is the internal coupling strength, $A_{ij}$ is the
adjacency matrix of internal connections; $k_i = \sum_j A_{ij}$ is the degree of node $i$; $F$ and
$\sigma$ are respectively the amplitude and frequency of the external force; and $C$ is the subgroup
of oscillators subjected to the external force. We have also defined $\delta_{i,C} = 1$ if $i \in C$
and zero otherwise and we shall call $N_C$ the number of oscillators in the set $C$. The natural
frequencies $\omega_i$ are taken from a distribution $\rho(\omega)$, which is here chosen to be
Gaussian with zero average.

Following \cite{Childs2008} we define
\begin{equation}
\phi_i = \theta_i - \sigma t.
\end{equation}
In these new variables the explicit time dependence disappears and the equations become
\begin{equation}
\dot{\phi_i} = \omega_i -\sigma - F \, \delta_{i,C} \sin \phi_i + 
\frac{1}{k_i} \sum_{j=1}^{N} \lambda_{ij} \sin (\phi_j - \phi_i),
\label{forced2}
\end{equation}
where $\lambda_{ij} = \lambda A_{ij}$. 

The adjacency matrix $A_{ij}$ gives the strength of interaction between oscillators $i$ and $j$. For
unweighted networks $A_{ij}$ assumed the value 1 if they interact and 0 otherwise, but weighted
networks like that of the {\it C. elegans}, might have very inhomogeneous distributions of weights.
For networks that can be divided into anatomical or functional communities, the external force can
be applied to one of the communities as a way to probe its influence on the others. Thus, we will
investigate how the control parameters, $\lambda$ and $F$, affect the spontaneous and induced
synchronization of the focal community (where the force is applied) and how it spreads to the other
communities of the system. 

If there is no external force and if  the internal coupling constant $\lambda$ is sufficiently large the oscillators
synchronize spontaneously with frequency $\bar\omega = \sum \omega_i/N$ in the original coordinates $\theta$ or with
frequency $\bar\omega - \Omega$ in the rotating frame $\phi$. On the other hand, if both $\lambda$ and $F$ are large the
system synchronizes with the external frequency $\Omega$ in the original frame or $-\bar{\omega}$ in the rotating frame.
In our simulations, since the Gaussian distribution is symmetric, $\bar{\omega}=0$, so that spontaneous synchronization
corresponds to global frequency $ \dot{\psi} =  - \Omega$ and forced synchronization to frequency $ \dot{\psi} =  0$.

Following \cite{us} we can estimate the minimum intensity of the external force, $F_c$, required to induce global synchronization using the relation 
\begin{equation}
F_{c} = \frac{\sigma}{f} \frac{\langle k \rangle}{\langle k \rangle_C},
\label{Fcrit}
\end{equation}
where $f=N_C/N$ is the fraction of forced neurons; $\langle k \rangle$ and $\langle k \rangle_C$ are  the average degree
of the network and the forced module, respectively.

\subsection{\textit{C. elegans} neural connectome}
\label{cenc}

Based on structural and functional properties of the neural network of \textit{C. elegans}, Varshney et al
\cite{chen2011} and Yan et al \cite{nature} presented a division of neuronal classes, totalizing 118, in three
categories: sensory neurons (SN), which respond to environmental variations, motoneurons (MN), recognized by  the
presence of neuromuscular junctions and responsible by locomotion, and the interneurons (IN), which cover all of other
classes. The adjacency weighted matrix is defined as follows: the element $w_{ij}$ represents the total number of
synapses interchange between the pair of neurons $ij$. In \cite{chen2011} the authors also divide the set into the gap
junction network, which refers to the electrical synapses, and  the chemical synapses network.

Gap junctions are a medium for electrical coupling between neurons and, since the electric signal can be made in both
directions, the electrical junction network is considered undirected and, consequently, its adjacency matrix symmetric.
On the other hand, the chemical synapses network is a directed and weighted network, whose adjacency matrix is
assymetric. Here we will concentrate on the electrical junction network only.

We analyzed the gap junctions neural network of nematode \textit{C. elegans} extracting the data from WormAtlas
\cite{wormatlas}. The full connectome has 279 neurons (nodes) and 514 gap junctions (connections) divided  into a giant
component with 248 neurons plus 31 neurons not connected with it. Here we will study the dynamics on the giant
component. Thereby, we built the weighted electrical junction (EJ) network of the \textit{C. elegans} with 248 neurons
and 511 gap junctions.  We also used a hierarchical algorithm to detect communities on the EJ network. For that, we used
the package ModuLand \cite{moduland} available on the free software Cytoscape. The algorithm provided three modules ($M_1$,
$M_2$, $M_3$) with modularity $Q = 0.44$.

% Figure 1
\begin{figure}[H]
\center
\includegraphics[scale=0.34]{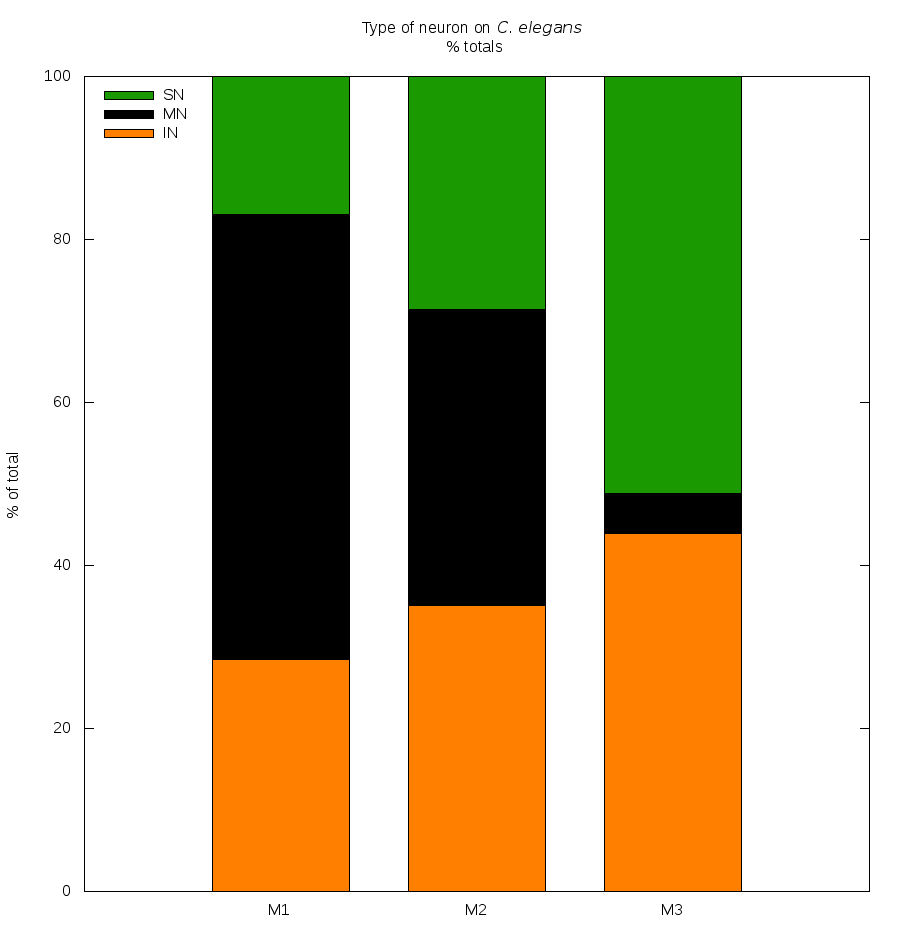}
\includegraphics[scale=0.34]{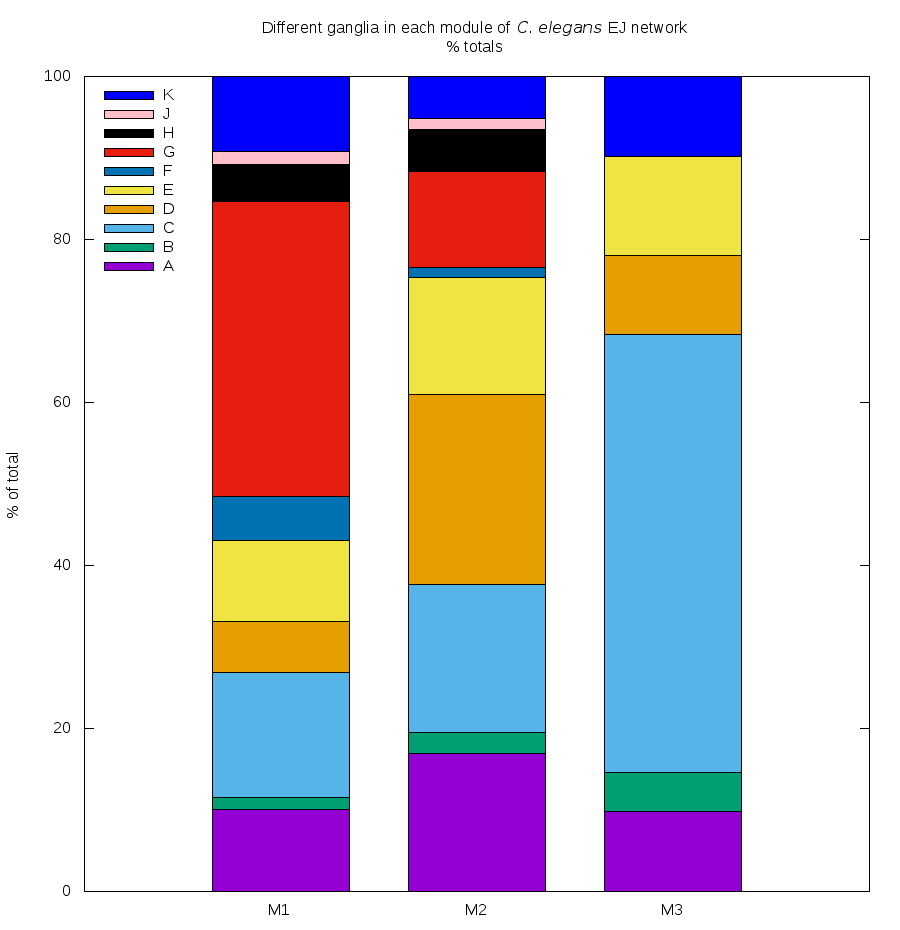}
\caption{(color online) Histograms representing the fraction of (left) neuronal class (SN: sensory neuron, MN: motorneuron 
and IN: interneuron) and (right) of ganglia (A: anterior ganglion, B: dorsal ganglion, C: lateral ganglion, D: ventral ganglion, 
E: retrovesicular ganglion, F: posterolateral ganglion, G: ventral cord neuron group, H: pre-anal ganglion, J: dorsorectal 
ganglion, K: lumbar ganglion) for each module ($M_1$: module 1, $M_2$: module 2 and $M_3$: module 3).}
\label{hist1}
\end{figure}

\begin{figure}[!htpb]
\center
\includegraphics[scale=0.34]{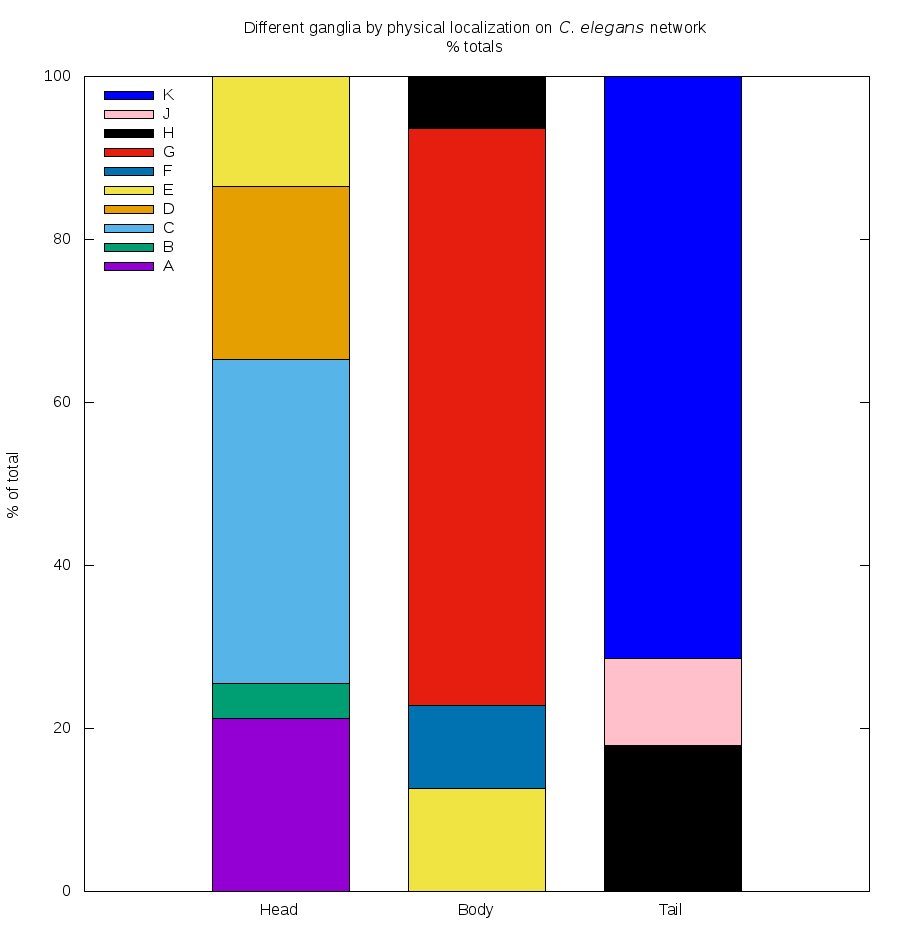}
\includegraphics[scale=0.34]{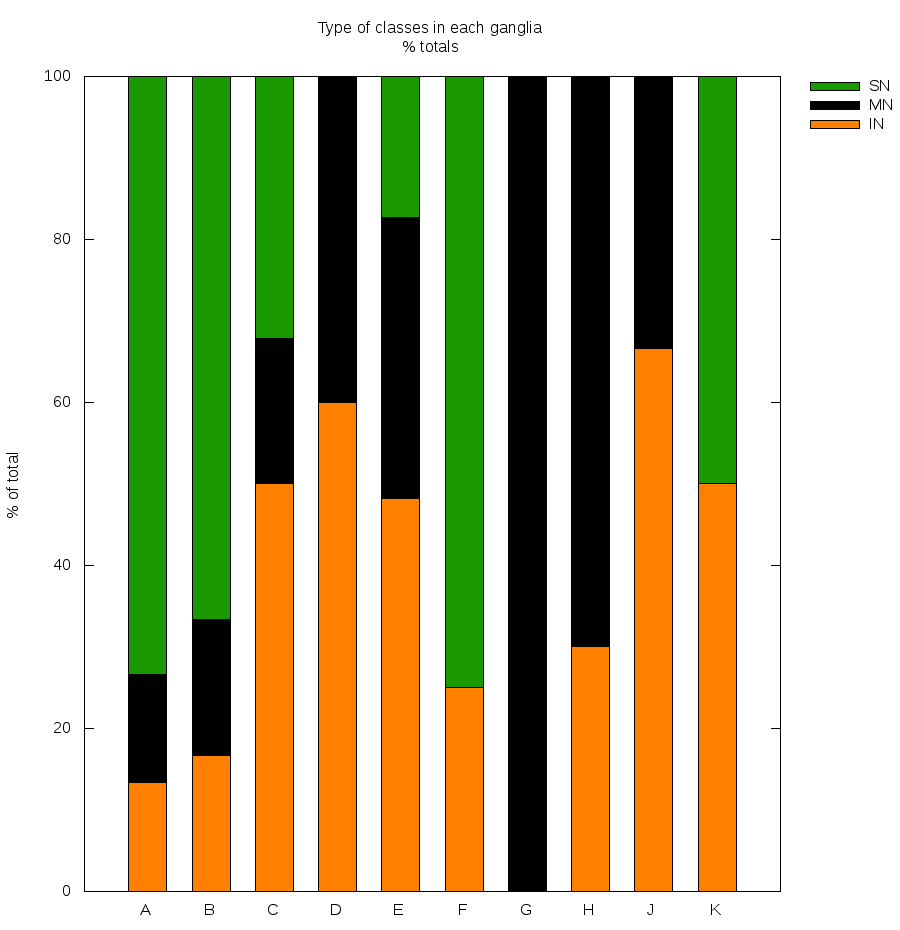}
\caption{(color online) Histograms representing the fraction of (left) different ganglia distributed by physical localization 
(head, midbody and tail) and (right) the fraction of classes (SN, MN and IN) component in each ganglia.}
\label{hist2}
\end{figure}

We classified each neuron as belonging to one of three categories (sensory, motor and interneurons) and based on ganglia
(A: anterior ganglion, B: dorsal ganglion, C: lateral ganglion, D: ventral ganglion, E: retrovesicular ganglion, F:
posterolateral ganglion, G: ventral cord neuron group, H: pre-anal ganglion, J: dorsorectal ganglion, K: lumbar ganglion
\cite{wormatlas}). The compositions of neuronal categories and ganglionic classification in each module are shown in
Figure \ref{hist1} (left) and \ref{hist1} (right), respectively. The ganglia are a finner division of the anatomical
classification into head (H), body (B) and tail (T). The left histogram of figure  \ref{hist2} shows that the
ganglia A, B, C and D belong to the head, G is entirely localized at the body, while J and K belong to the tail. 

% Figure 2

This set of divisions of the neural network into communities can be classified as (i) topological ($M_1$, $M_2$, $M_3$); (ii)
anatomical (ganglia) and; (iii) functional (SN, MN, IN). They are clearly all different, showing that EJ network has a
complex modular structure. Table \ref{table3} shows the relevant data to modular network used. We calculated the fraction $f$ and the average degree $\langle k \rangle_C$ of each classification and of the whole network, $\langle k \rangle$, and the respective values of critical force as given by Eq.(\ref{Fcrit}).

\begin{table}[H]
\center
\caption{Number of neurons, fractions of nodes, average degree and theoretical critical force for each classification and for the whole network.}
\begin{tabular}{|c|c|c|c|c|}
\hline
\multicolumn{1}{|l|}{} & Module 1                                           & Ganglia C                                           & Sensory Neurons                                           & Network                                          \\ \hline
Number of neurons      & $N_{M_1}$ = 130                                        & $N_C$ = 56                                          & $N_{SN}$ = 65                                         & $N$ = 248                                              \\ \hline
Fraction of nodes      & $f_{M_1}$ = 52.42 \%                                     & $f_C$ =  22.58 \%                                     & $f_{SN}$ =26.21 \%                                       & $f$ = 100 \%                                           \\ \hline
Average degree         & $\langle k_{M_1} \rangle $ = 7,96 & $\langle k_C \rangle $ = 10,16 & $\langle k_{SN} \rangle $ = 5,27 & $\langle k \rangle $ = 7,13 \\ \hline
Critical force (theoretical)   & \multicolumn{1}{l|}{$F_{c,theo}^{M_1} $ = 5,17} & \multicolumn{1}{l|}{$F_{c,theo}^{C} $ = 9,32} & \multicolumn{1}{l|}{$F_{c,theo}^{SN} $ = 22,98} & \multicolumn{1}{l|}{$F_{c,theo} $ = 3,00} \\ \hline
\end{tabular}
\label{table3}
\end{table}

\subsection{Order parameters and correlations}
\label{op}

The partially forced Kuramoto dynamics will be applied to the {\it C. elegans} as a way to probe its
modular structure. Forcing a particular module may or may not induce synchronization with the
external frequency on other modules of the system. In order to monitor the behavior of separate
modules we define 
\begin{equation}
z_n = \frac{1}{N_n} \sum_{i \in M_n} e^{i \phi_i} \equiv r_n e^{i \psi_n}
\end{equation}
where the subscript $n$ specifies the module $M_n$ of size $N_n$. Therefore, $r_n$ is a local order parameter that
measures how much the oscillators in the module are synchronized among themselves. The angular velocity $\dot{\psi_n}$
provides information about the motion of the set: $\dot{\psi_n} = 0$ implies sync with the external force, $\dot{\psi_n}
= -\Omega$ refers to spontaneous sync whereas  nonconstant values indicate more complex behavior.

Intermodule behavior will also be monitored by the quantities
\begin{equation}
z_{nm} = \frac{1}{N_n+N_m} \sum_{i \in M_n \cup M_m} e^{i \phi_i} \equiv r_{nm} e^{i \psi_{nm}}
\end{equation}
with similar interpretations.

We also compute velocity-velocity correlations between all pairs of oscillators. We define
\begin{equation}
\tilde{c}(i,j) = \frac{1}{T} \int_{t_0}^{t_0+T} (\dot{\phi}_i(t) - \langle \dot{\phi_i} \rangle)
(\dot{\phi}_j(t) - \langle \dot{\phi_j} \rangle) \, dt
\label{fluc1}
\end{equation}
where
\begin{equation}
\langle \dot{\phi_i} \rangle = \frac{1}{T} \int_{t_0}^{t_0+T} \dot{\phi}_i(t)  \, dt
\end{equation}
and $t_0$ is a sufficiently long time so that the transient dynamics has passed. 

The normalized velocity-velocity correlation function is then defined as:
\begin{equation}
c(i,j) = \frac{\tilde{c}(i,j)}{\sqrt{\tilde{c}(i,i) \, \tilde{c}(j,j)}},
\label{corr-v-v}
\end{equation}
where $|c(i,j)| \leq 1 $. We note that the correlation is computed in terms of the fluctuations of the average velocity,
that was subtracted out in Eq. (\ref{fluc1}). The 248 $\times$ 248 correlation matrix gives direct information about the
effect of one neuron over another, irrespective of their synchronization state. If an increase in the velocity of $i$
leads to the average increase in the velocity of $j$ then nodes $i$ and $j$ are positively correlated and $c(i,j) > 0$.
If, on the other hand the velocity of $j$ decreases, they are negatively correlated and $c(i,j) < 0$. Finally, if they
are uncorrelated $c(i,j) \approx 0$. In the simulations we used $t_0=T/2$ and $T=20$ which was enough
for the equilibration of the system. 

The parameters $z_n$ provide information about the synchronization of each module, whereas the average value of
the phase velocity $\dot{\psi}_n$ tells whether the module follows the external force or  spontaneous collective
motion. This information is complemented by the velocity-velocity correlation, which measures the effect of one node over the
other even if they synchronize with different frequencies or are not synchronized at all.

% % % % % % % % % % % % % % % % % % % % % % % % % % % % % % % % % % %
% % % % % % % % % % % % % % % % % % % % % % % % % % % % % % % % % % %
 \newpage
\section{Results}
\label{results}

\begin{figure}[H]
\center
\includegraphics[scale=0.24]{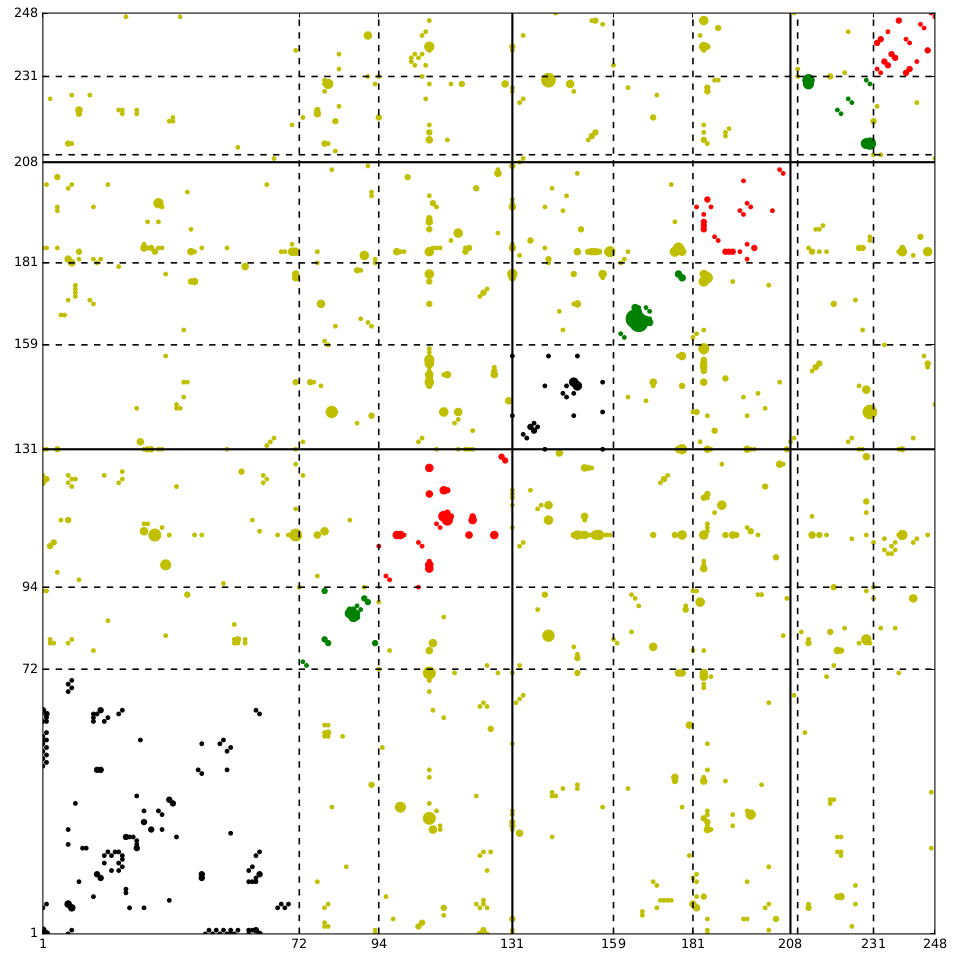}
\includegraphics[scale=0.24]{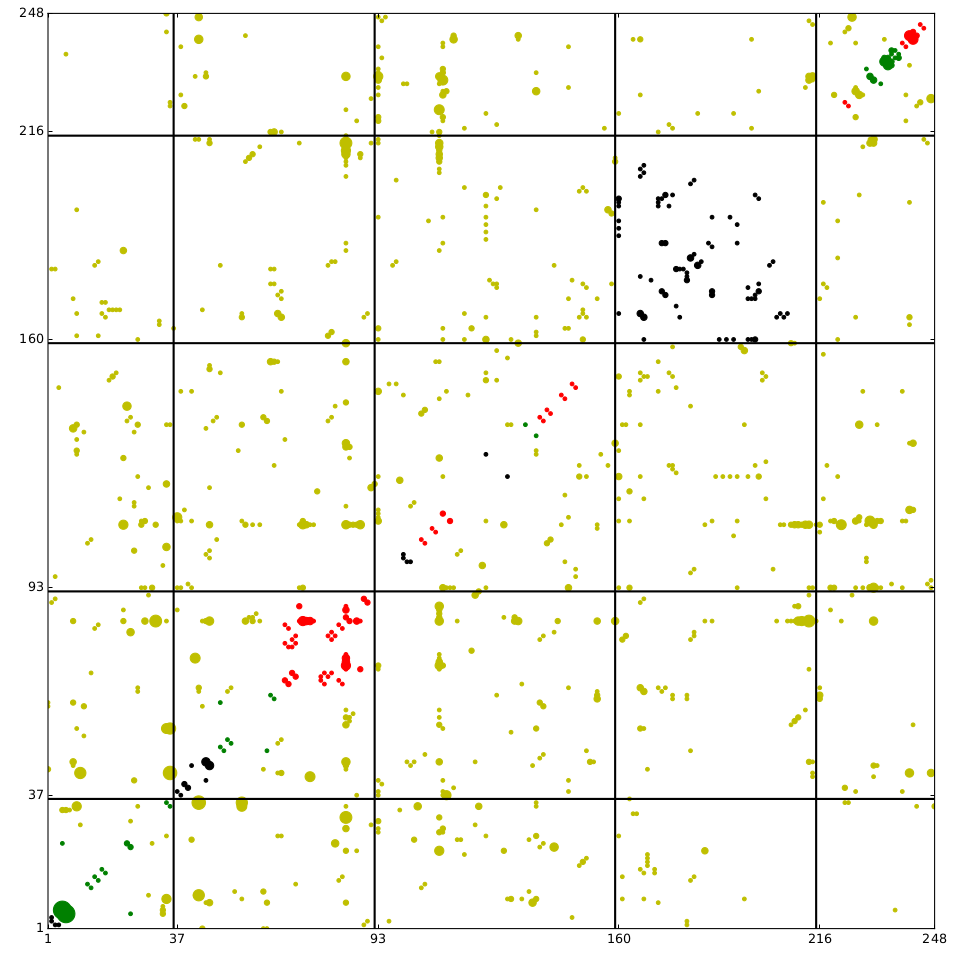}
\caption{(color online) Left panel: weighted adjacency matrix highlighting the 3 topological modules $M_1$, $M_2$ and $M_3$ separated 
by thick black lines and subdivided into motor (black), sensory (green) and interneurons (red) by dashed black lines. Intermodule connections are shown in yellow. Right panel: weighted adjacency matrix highlighting 5 groups of ganglia separated by thick 
black lines, \{A, B\}; \{C\}; \{D, E, F\}; \{G\} and \{H, J, K\}. Subdivisions and intermodule connections follow the functional colors of left panel.}
\label{adjs}
\end{figure}

Figure \ref{adjs}  shows the weighted adjacency matrix ordered according to the topological modules $M_1$, $M_2$ and $M_3$ (left panel).
Modules are separated by thick black lines and subdivided into motoneurons (black), sensory neurons (green) and
interneurons (red) by dashed black lines. The size of the dot is proportional to the intensity $A_{ij}$ and intermodule
connections are represented in yellow. The right panel shows the adjacency matrix ordered by ganglia, from head to tail.
The thick black lines highlight 5 groups of ganglia: \{A, B\}; \{C\}; \{D, E, F\}; \{G\} and \{H, J, K\}. Subdivisions and intermodule
connections follow the functional colors of left panel. The indexes in both panel delimit the divisions made.

% Figure 3

In order to analyze the interdependencies of the modules for  different partitions of the EJ network, we have simulated
the application of an external stimulus to one of the modules and observed its effect on the others. The stimulus is
modeled by an external periodic force acting only on the selected module under the Kuramoto dynamics as described in
section \ref{pfk}. In all our simulations we have fixed $\Omega = 3$. The effects on the other modules
is measured by local order parameters, such as $r_n$, and the normalized velocity-velocity correlation function as
described in section \ref{op}. The numerical results are shown in figures 4, 5 and 6 below and in figures 1, 2 and 3 of
SM. In what follows we show results when the stimulus is applied only to $M_1$, or to ganglion C or to the sensory
neurons.\\

\noindent  \textbf{Stimulating the $M_1$: the role of topology} \\

%\noindent {\it Module 1} 

\begin{figure}[!htpb]
\center
\includegraphics[scale=0.46]{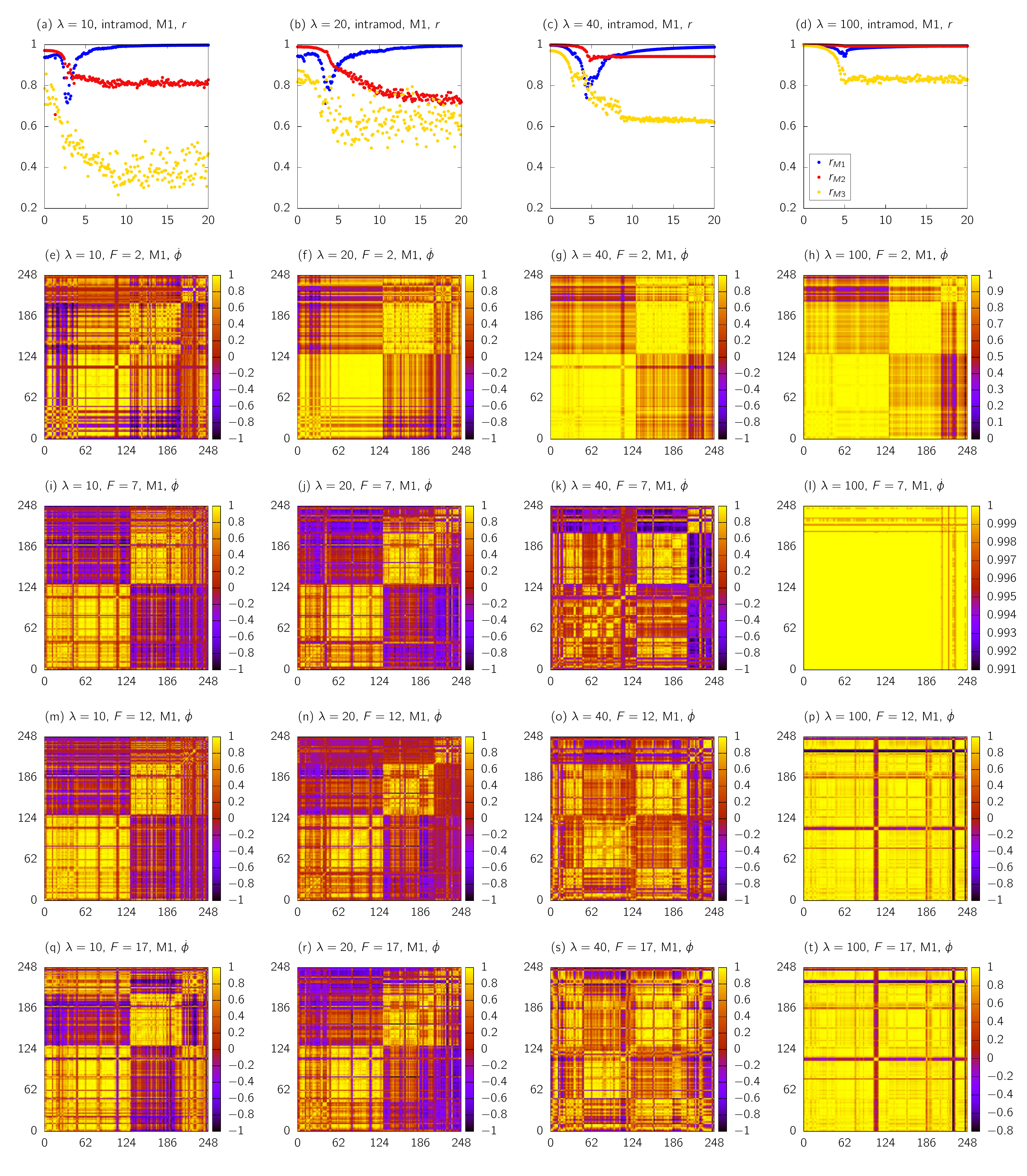}
\caption{(color online). Panels (a)-(d): the local order parameters as a function of the external force $F$ acting on neurons of $M_1$ for $\lambda$ fixed. Panels (e)-(t): the velocity-velocity correlation matrix 248 $\times$ 248 obtained using Eq. (\ref{corr-v-v}). In each panel, the fixed parameters $\lambda$ and $F$ are indicated. The $M_1$ neurons are indexed by 1 to 130, the $M_2$ neurons by 131 to 207 and the $M_3$ neurons are indexed by 208 to 248.}
\label{corr-M1-r}
\end{figure}

Figure \ref{corr-M1-r} shows the results of simulations for four values of the internal coupling $\lambda$ when only the
neurons of $M_1$ are forced (indexes 1 to 130 in the left panel of figure \ref{adjs}.) The local order parameters $r_n$
as a function of the intensity of the external force $F$ are exhibited in panels (a) to (d) and the velocity-velocity
correlation matrices are in panels (e) to (t). The neurons of $M_1$ are spontaneously synchronized for $F=0$ with
$\dot\psi_{M_1} = - \Omega$ (see Fig. 1 on SM). As $F$ increases, the neurons go through a region of desynchronization
around $F=5$, which is close to theoretical value $F_{c,theo}^{M_1} $ = 5.17 calculated from eq. (\ref{Fcrit}), and then
the neurons synchronize with the external force ($\dot{\psi}_{M_1}=0$, Fig. 1 on SM) for $F$ larger than about 10, where
$r_{M_1} \rightarrow 1$. For large internal coupling $\lambda$, all modules sync with external force (see panel (d) of
Fig. \ref{corr-M1-r} and panel (l) of Fig. 1 on SM), but $M_3$ has large fluctuations in $\dot\psi_{M_3}$ (panel (p),
Fig. 1 on SM). Besides, $M_3$ does not fully synchronize since the order parameter reaches its maximum value for
$r_{M_3} = 0.8$.

The most striking feature of these simulations is the strong anti-correlation patterns developed between $M_1$
and $M_2$ for $\lambda \leq 20$. From the top panels we notice that, in these cases, $M_1$ is synced with the external
force and $M_2$ is synchronized spontaneously. Nevertheless the effects of $M_1$ over $M_2$ are very clearly shown by
the purple areas of the correlation plots. \\

%%%%%%%%%%%%%%%%%%%%%%%%%%%%%%%%%%%%%%%%%%%%%%%%%%%%%%%%%%%%%%%%%%%%%%%%%%%%%%%%%%%%%%%%%%%%%%%%%%%%%%%%%%%%%%%%%%
\noindent  \textbf{Stimulating ganglion C: the role of anatomy} \\

%\noindent {\it Ganglion C}
  
\begin{figure}[!htpb]
\center
\includegraphics[scale=0.46]{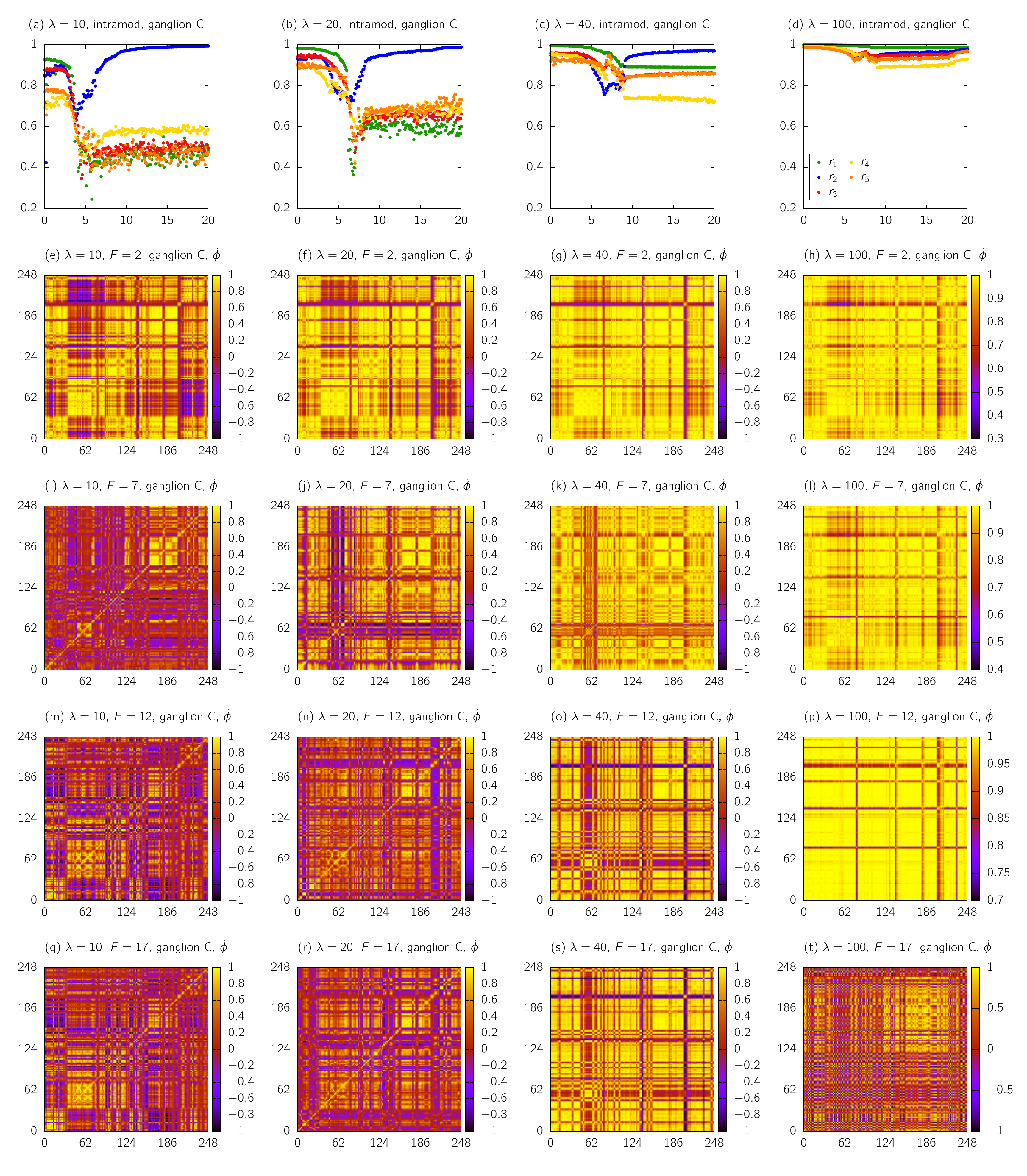}
\caption{(color online). Panels (a)-(d): the local order parameters as a function of the external force $F$ acting on ganglion C for $\lambda$ fixed. Panels (e)-(t): the velocity-velocity correlation matrix 248 $\times$ 248 obtained using Eq. (\ref{corr-v-v}). In each panel, the fixed parameters $\lambda$ and $F$ are indicated. The group 1 (\{A,B\}) are indexed by 1 to 36, group 2 (\{C\}) by 37 to 92, group 3 (\{D, E, F\}) by 93 to 159, group 4 (\{G\}) by 160 to 215 and group 5 (\{H, J, K\}) are indexed by 216 to 248.}
\label{corr-C}
\end{figure}

Figures \ref{corr-C} shows the results of simulations when ganglion C is forced. The values of the internal coupling
$\lambda$ are the same used in Fig. \ref{corr-M1-r}. The local order parameters $r_n$ as a function of $F$ are in panel
(a) to (d) of each figure. The behavior of the order parameters $r_n$ as a function of $F$ is similar to that observed
when forcing the neurons of $M_1$, exhibiting a region of desynchronization between $F=5$ and $F=10$, which contains the
theoretical value calculated by eq. (\ref{Fcrit}), $F_{c,theo}^C = 9.32$, followed by stabilization for larger $F$. The
forced neurons are clearly seen as a bright yellow block in panels (e) to (h).

For sufficiently large $F$ ganglion C synchronizes with the external force ($r_C \rightarrow 1$, panels (a) to (d), and
$\dot{\psi}_{C}=0$, Fig. 2 on SM) for all values of $\lambda$ considered. However, the velocity-velocity correlation
matrices show much simpler patterns: either nearly complete correlation occurs
(yellow areas in panels (g), (h), (k), (l) and (p)),  or almost no correlation is observed (large red areas in panels
(i), (m), (n), (q) and (r)) where ganglion C itself shows reduced internal correlations. Even for  $F>12$, where $r_C$
indicates that C is nearly fully synchronized for all $\lambda$'s, the correlation matrices show regions of mixed
behavior, especially for small $\lambda$, which means that part of neurons of C are non-correlated with each
	other or even anti-correclated (see also Fig. 2, panels (a), (b), (i) and (j) on SM). Although all ganglia seem to synchronize with the
external force for $\lambda \geq 40$ and $F > 12$, their dynamics are uncorrelated with other ganglia.  The
only exception is ganglia G, that shows up as a yellow square in the plots  (see also Figs. 4 and 5 on SM).

We also note that for $\lambda = 40$ the motor part of ganglion C (small yellow squares indexed by 37 to 46) correlates
separately from the rest of C for $F=12$, panel (o), and $F=17$, panel (s), which means that motoneurons respond
differently to external stimuli. For $\lambda = 20$ fixed, the motoneurons of ganglion C first synchronize spontaneously
($F=2$ panel (f) and Fig. 2, panels (b) and (f) on SM), then exhibit mixed behavior, with most pairs 
	uncorrelated  ($F=7$ panel (j) and Fig. 2, panels (b) and (j) on SM). As the force intensity increases the group
regains part of the synchrony (now with the external force) but never fully correlated (panels (k) and (l) and regions
of $F>10$ in panels (b), (c) and (d)).
  
For $F=17$, panels (q), (r), (s) and (t), the number of correlated neurons increases from $\lambda=10$ to $\lambda=40$
but for $\lambda=100$ the entire network goes out of phase, with the exception of ganglion G: it keeps its internal
correlation at all times, maybe because it is entirely a motor ganglion type. Note that $r_G$ is close to 1 only for
$\lambda = 100$, panel (d), which means that full synchronization requires large internal coupling. \\

\noindent  \textbf{Stimulating the sensory neurons: the role of function} \\

\begin{figure}[!htpb]
\center
\includegraphics[scale=0.46]{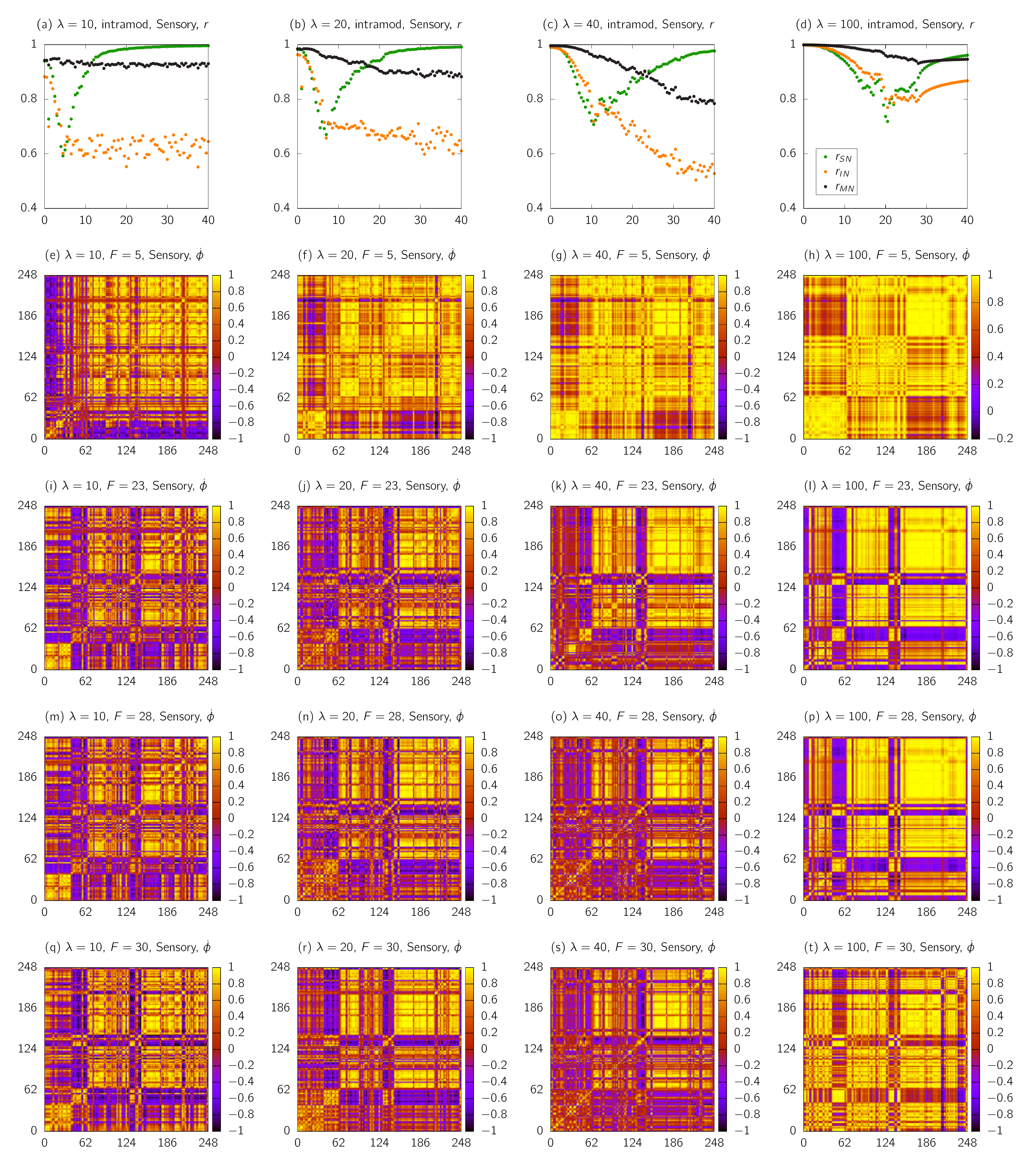}
\caption{(color online). Panels (a)-(d): the local order parameters as a function of the external force $F$ acting on sensory neurons for $\lambda$ fixed. Panels (e)-(t): the velocity-velocity correlation matrix 248 $\times$ 248 obtained using Eq. (\ref{corr-v-v}). In each panel, the fixed parameters $\lambda$ and $F$ are indicated. The sensory neurons (SN) are indexed by 1 to 65, the interneurons (IN) by 66 to 147 and the motoneurons (MN) are indexed by 148 to 248.}
\label{corr-sensory}
\end{figure}

Figure \ref{corr-sensory} shows the numerical results when all sensory neurons receive the external stimuli. In this
case, we divided the network in three groups based on functional classes - sensory neurons, interneurons and
motoneurons. The values of internal coupling are the same used previously. From panels (a) to (d) it can be seen that
the behavior of sensory neurons is complex, synchronizing with the external force for: $i)$ $\lambda = 10$ and $F>10$ on
panels (a) and (i) of Fig. 3 (see SM), $ii)$ $\lambda = 20$ and $F>15$ on panels (b) and (j) of Fig. 3 (see SM), $iii)$
$\lambda = 40$ and $F>20$ on panels (c) and (k) of Fig. 3 (see SM) and $iv)$ $\lambda = 100$ and $F> 30$ on panels (d)
and (l) of Fig. 3 (see SM). In $iii)$ and $iv)$ the values of $F$ are close to the theoretically calculated one, which
is $F_{c,theo}^{SN} = 22.98$. Differently from all other cases, larger values of $\lambda$
hinders the synchronization of the forced group, since $r_{SN}$ decreases from $\lambda=10$ to $\lambda=100$,
although $\dot{\psi_{SN}} = 0$ (Fig. 3, SM).

For $\lambda \leq 40$, the motoneurons and almost half of interneurons were spontaneous sync (panels (a), (b), (i) and
(j) on Fig. 3, SM), while for $\lambda > 40$ and $F>30$ most neurons were synchronized with external stimuli.
Furthermore, the results of velocity-velocity matrices also show regions of anti-correlation and non-correlation, as can
be seen on purple and red areas of Fig. \ref{corr-sensory}, respectively, and they are more prominent for weak internal
coupling, $\lambda < 40$. In these cases, the lack of correlation seems to indicate a lower value of
inter-modules  order parameter $r$, as can be seen in panels (e) to (h) of Fig. 3 on SM.

% Figure 4

% Figure 5

% Figure 6

%%%%%%%%%%%%%%%%%%%%%%%%%%%%%%%%%%%%%%%%%%%%%%%%%%%%%%%%%
\newpage
\section{Discussion}
\label{discussion}

Information processing in the brain requires the synchronous firing of specific groups of neurons as a response to
external stimuli. Groups of neurons can be defined in many ways, taking into account their anatomical location, their functional role or their topological properties in the network. In this paper we have
investigated the importance of these divisions as targets to stimuli, as well as their roles in spreading the inputs to
other parts of the brain. We used a much simplified model of synchronization given by the Kuramoto system of phase
oscillators subjected to localized external stimuli that was applied to a subset of neurons representing a
topological module (Fig. \ref{corr-M1-r}), a ganglion composed of different functional neurons (Fig. \ref{corr-C}) or
all sensory neurons (Fig. \ref{corr-sensory}). The response of  the electrical neural network to the stimulus was
different is each case, as we summarize below.

%\textbf{About topology:}
The modularization procedure applied to EJ network reveals that the modules do not contain purely anatomical groups or functional classes, but mixes neurons belonging to different ganglia and functional classes, such as sensory, motor and interneurons. This is illustrated in Fig. \ref{hist1}, where we have analyzed the distribution of neuronal class and ganglia membership in each module. This corroborates previous studies \cite{antono2015, pan2010, sohn, chen2011,arenas2007} and shows the complexity of the  neuronal wiring and function.

Stimulation of the neurons of the largest topological module $M_1$ induced strong anti-correlation in the velocity
fluctuations of the neurons in $M_2$ and $M_3$ (purple areas of panels (i), (j) and (m) on Fig. \ref{corr-M1-r}
or between $M_2$ and $M_3$ (panel (k) on the same figure), which kept their original state of spontaneous
synchronization for moderate values of the internal coupling constant $\lambda$. This anti-correlation also seems to
suggest a lower value of the inter-modules order parameter $r$, as can be seen between $M_1$-$M_2$ and $M_2$-$M_3$
(panel (k) on Fig. \ref{corr-M1-r} and Fig. 1 on SM). On the other hand, the presence of positive
correlations between $M_1$ and $M_3$ (panel (q) on Fig. \ref{corr-M1-r}), is accompanied by an increase of $r_{13}$.

The smallest topological module $M_3$ remained oblivious to the stimulus even for large values of $\lambda$.
Interestingly, for intermediate values of the forcing (panel (j) on Fig. \ref{corr-M1-r}), the neurons of $M_1$ became
mostly uncorrelated (red areas on Fig. \ref{corr-M1-r}), indicating a parameter region of poor response to the
stimulus. It is possible, however, to identify the modular structure by the presence of three blocks, each of
one corresponding to $M_1$, $M_2$ and $M_3$.

%\textbf{About ganglion:}
Ganglia are structures formed by the grouping of cell bodies of neurons and are responsible for the control of specific
regions. In \textit{C. elegans} it was possible to classify 10 ganglia groups, since most of the cell bodies of neurons
are clustered in these structures \cite{wormatlas}. The response of the network to stimulation of ganglion C was quite
different from the other two cases we considered, displaying essentially two distinct regions with
(I) large parameter intervals of almost complete uncorrelated behavior, which occurs for $\lambda \leq 20$
(red areas on panels (i), (m), (n), (q), (r) and (t) on Fig. \ref{corr-C}) and (II) complete correlated
behavior, with $\lambda \geq 40$ (yellow areas on Fig. \ref{corr-C}). Effective synchronization of ganglion C with the
external force required large values of the coupling constant. Contrary to what occurs when forcing the
topological module, the blocks of the correlation matrix corresponding to ganglia groups
cannot be clearly distinguished, except for ganglion G (Figs. 4 and 5 on SM), which seems to hold high correlation
between its neurons, possibly because it is the only group entirely composed by one class (motoneurons).

%\textbf{About sensory neurons:}  

Sensory neurons are responsible for collecting information from external environment and react to stimuli inside the
organism, acting as an input channel. In this sense, \textit{C. elegans} uses these neuronal functions to explore the
ambient, navigating over thermal, chemical and oxygen variations, in addition to avoid hostile behavior
\cite{wormatlas}. The response of these neurons to the stimulus leads to induced synchronization for $\lambda
\leq 40$, while the other two classes show spontaneous sync. It was only with strong internal coupling, $\lambda > 40$,
and force larger than the theoretical value, $F>30$, that most of neurons were induced to the forced sync. The results
also show many regions of anti and non-correlation (purple and red areas on Fig. \ref{corr-sensory}, respectively) and
blocks relative to the three classes are reasonably clear, although showing visible internal structure, which indicates
a more complex relationship between them.

%\textbf{About our model and everything}
The values of the critical force estimated theoretically with Eq. (\ref{Fcrit}) are close to numerical results found by stimulating the EJ network of nematode \textit{C. elegans} in all cases studied. In particular, for $F > F_{c,theo}$ and strong coupling $\lambda$ the
stimulated group syncs with $F$ in all cases. We also found that synchronization and positive correlation between
modules seem to be related, since yellow areas on the correlation matrices occurs when $r \rightarrow 1$.

Previous studies \cite{us} showed that the Kuramoto model with external localized stimuli leads to global
synchronization on synthetic networks if $\lambda$ and $F$ are sufficiently large. Here we considered a real neural
network and we did not observe full synchronization, which indicates that the particular modular structure of the
network protects the system from `seizures'. The group divisions considered here (topological, functional and
anatomical) are natural but not complete and finer subdivisions might be important understand the system response in more detail.
This shows the complex structure of the neural network. \\ \\

\noindent Acknowledgments: M.A.M.A.  acknowledges financial support from CNPq (grant 302049/2015-0) and
FAPESP (grants 2016/06054-3 and 2015/11985-3). C.A.M. was supported by CNPq (grant 141021/2017-9).

%%%%%%%%%%%%%%%%%%%%%%%%%%%%%%%%%%%%%%%%%%%%%%%%%%%%%%%%%
\clearpage 
\newpage
%\bibliographystyle{ieeetr}
%\bibliography{kuramoto}

\end{document}